\documentclass[a4paper]{article}

\usepackage[english]{babel}
\usepackage[utf8]{inputenc}
\usepackage[T1]{fontenc}
\usepackage{amsmath}
\usepackage{amsfonts}
\usepackage{graphicx}
\usepackage{array}

\makeatletter
\def\underbracket{%
\@ifnextchar[{\@underbracket}{\@underbracket [\@bracketheight]}%
}
\def\@underbracket[#1]{%
\@ifnextchar[{\@under@bracket[#1]}{\@under@bracket[#1][0.4em]}%
}
\def\@under@bracket[#1][#2]#3{
\mathop{\vtop{\m@th \ialign {##\crcr $\hfil \displaystyle {#3}\hfil $%
\crcr \noalign {\kern 3\p@ \nointerlineskip }\upbracketfill {#1}{#2}
\crcr \noalign {\kern 3\p@ }}}}\limits}
\def\upbracketfill#1#2{$\m@th \setbox \z@ \hbox {$\braceld$}
\edef\@bracketheight{\the\ht\z@}\bracketend{#1}{#2}
\leaders \vrule \@height #1 \@depth \z@ \hfill
\leaders \vrule \@height #1 \@depth \z@ \hfill \bracketend{#1}{#2}$}
\def\bracketend#1#2{\vrule height #2 width #1\relax}
\makeatother

\title{Coexistence of several currencies in presence of increasing returns to adoption}

\author{Alex Lamarche-Perrin$^{(1)}$, André Orléan$^{(2)}$, Pablo Jensen$^{(1)}$\\
  \multicolumn{1}{p{.99\textwidth}}{\centering\emph{$^{(1)}$ Univ Lyon, Ens de Lyon, Univ Claude Bernard, IXXI, CNRS, Laboratoire de Physique, F-69342 Lyon, France \\ $^{(2)}$ CNRS EHESS ENPC ENS, UMR 8545, Paris-Jourdan Sciences Economiques (PjSE)
48 Boulevard Jourdan, 75014 Paris, France}}}

\begin{document}

\maketitle

\begin{abstract}
We present a simplistic model of the competition between different currencies. Each individual is free to choose the currency that minimizes his transaction costs, which arise whenever his exchanging relations have chosen a different currency. We show that competition between currencies does not necessarily converge to the emergence of a single currency. For large systems, we prove that two distinct communities using different currencies in the initial state will remain forever in this fractionalized state.
\end{abstract}



\section{Introduction}
In general, a currency is useful - and therefore sought by an individual - only insofar as it can be used to buy goods. This implies that the currency is widely accepted as payment by her suppliers. In a world where several currencies exist, the attractiveness of a currency for an individual can be measured by its greater or less acceptance in the group of individuals with whom it is used to exchange goods. It follows that the more a currency is accepted, the greater its attractiveness becomes. From this point of view, money is akin to language  \cite{orlean}. The assumption that in cross-currency competition, increasing returns to adoption play a fundamental role has been present in economic theory since the famous article by Carl Menger \cite{menger}. The same idea has been examined by various authors in the Economics litterature \cite{jones,kiyotaki,greenaway,sethi,duffy,lapavitsas,kunigami,gangotena} as well as in Physics journals \cite{yasutomi,gorski10,gorski15}. Recent models have pushed further Menger's basic idea either by trying to introduce social aspects of money \cite{lapavitsas}, or providing a unified framework able to explain at the same time the emergence of a single currency and some other economic phenomena. Yasutomi's\cite{yasutomi} model, further refined by \cite{gorski10,gorski15}, can also describe the collapse of a currency, while Donangelo and Sneppen's links the emergence of money to its anomalous fluctuation in value \cite{sneppen}. Duffy and Ochs \cite{duffy} tested some predictions of Kiyotaki and Wright model\cite{kiyotaki} in laboratory experiments. Another stream of literature connects the emergence of money to the more general question of the emergence of social norms using game theory\cite{greenaway,duffy,sethi,gangotena}. 

In this article, we propose a model that differs from all these approaches by two characteristics : its exchange mechanism is simpler and it takes into account the specificity of local situations through the introduction of a network of exogenously fixed bilateral links, for example because of constraints originating in the social division of labor. On the first point, most previous work assume some more  realistic exchange mechanism. For example, models inspired by Yasutomi's model \cite{yasutomi,gorski10,gorski15} take, as elementary interaction between two agents, the ”transaction”, which "consists of several steps including search of the co-trader, exchange of particular goods, change of the agent’s buying preferences and finally the production and consumption phase". On the second point, unlike most models, we do not assume a completely connected interaction network, which leads to simpler analytical treatments but obscures the local aspects of economic transactions. We demonstrate that, under such conditions, competition between currencies does not necessarily converge to the emergence of a single currency. Even if an equilibrium with a single currency remains possible, the most frequent stable configuration is the division of the trading space between different currencies. For large systems, we prove that two distinct communities using different currencies in the initial state will remain forever in this fractionalized state.

\section{General framework}

We consider an economy composed of $N$ agents, numbered $i = 1, ..., N $, each starting with its own currency $s_i$. A currency will be referred to as an integer in $[1,N]$, and we assume that each agent begins with its own currency $s_i = i$. The agents are disposed on the vertices of a random graph \cite{RG}, whose edges represent commercial links between the agents. After choosing a density of links $p \in [0,1]$, for each pair of agents $(i,j)$, we create a link between $i$ and $j$ with probability $p$, or let the agents disconnected with probability $(1 - p)$.

The interaction dynamics is set in the following way. Each time step, we choose an agent at random. First, this agent is allowed to change the currency it uses. Then, it trades with all its neighbours, {\emph i.e.} with all the agents he shares a link with. The profit an agent gets is defined according to the following idea: if two agents share the same currency, their trading business is made easier and no cost has to be paid; conversely, if they use different currencies, they must trade through the help of some "moneychanger" who gets a commission for its work: we will then consider that those agents have to afford some fixed cost (which we will take as a unit cost) in order to complete their trade. As there is no other constraint, we translate the profit of each particular trade to the origin so that a successful trade is worth $0$ and a trade which has to resort to a moneychanger is worth $-1$.

We define a simple utility function $U_i(t)$ for an agent $i$ at period $t$, as the opposite of the sum of all trading costs an agent has to pay when realising its trades at period $t$:
\[U_i = - \sum_{j\in \mathcal V(i)}(1 - \delta_{s_i}^{s_j}) = - Card\big\{j\in \mathcal V(i) | s_j \neq s_i\big\}\]
where $\mathcal V(i)$ is the set of the neighbours of the agent $i$ in the graph and $\delta_{s_i}^{s_j}$ is the Kronecker symbol, ie  
$\delta_{s_i}^{s_j} = 
\left \{
\begin{array}{c}
1 \text{ if } s_i = s_j \\
0 \text{ if }s_i \neq s_j
\end{array}\right.$.

We also define a social utility function for the economy as a whole, as the sum of the utilities of all agents:
\[\mathcal U = \sum_{i=1}^N U_i = - \frac 12\sum_{i,j\text{ neighbours}} (1 - \delta_{s_i}^{s_j})\]
where the factor $\frac 12$ accounts for the fact that each pair of neighbours is counted twice in the sum.

As we assume that agents are fully rational and maximize their own utility function, the rule for currency adoption is the simplest mimetic one: an agent adopts the most common currency among its neighbours; if several currencies are used by the same maximal number of neighbours, two cases appear: either the agent already uses one of them, and keeps using it by default, or she was using another one, in which case she picks at random one of the most popular currencies among her neighbours. Agents do not take into account neither the anticipated cost of future trades (for $t' > t$) nor the influence their choice could have on other agents.


From the evolution rule we have defined, we can infer that social utility can only increase with time; more, it increases strictly when an agent changes its currency. Indeed, if agent $i$ switches from currency $s$ to currency $s'$ at time $t$, social utility increases with
\begin{eqnarray}
\Delta \mathcal U 	& =	& \Delta U_i + \sum_{j \in \mathcal V(i) } \Delta U_j \nonumber \\
				& =	& \sum_{j \in \mathcal V(i) }  \delta_{s'}^{s_j} - \delta_s^{s_j} + \sum_{j \in \mathcal V(i) } \delta_{s_j}^{s' } - \delta_{s_j}^s \nonumber\\
				& = 	& 2 \bigg(Card\big\{j\in\mathcal V(i)|s_j = s'\big\} - Card\big\{j\in\mathcal V(i)|s_j = s\big\} \bigg)\nonumber
\end{eqnarray}
where we write $s_j = s_j(t-1) = s_j(t)$ for all $j \in \mathcal V(i)$. By definition, agent $i$ switches from $s$ to $s'$ at time $t$ if and only if 
\[Card\big\{j\in\mathcal V(i)|s_j = s'\big\} > Card\big\{j\in\mathcal V(i)|s_j = s\big\}, \]
 which yields 
\begin{equation}
\label{utiliteaugmente}
\Delta \mathcal U > 0.
\end{equation}

An interesting consequence of equation (\ref{utiliteaugmente}) is that the system can not come back to a state it has already visited: There exists no loop in the phase diagram. Because the number of states the system can visit is finite, we can infer that from any initial configuration, the system will reach an equilibrium with probability $1$ as time tends to infinity, equilibrium being defined as any state in which no agent can change its currency.  Rather than studying the precise dynamics of the system, we will hence be more interested in finding the different equilibria that this economy could reach. 

\begin{figure}
\begin{center}
\includegraphics[width = 4cm]{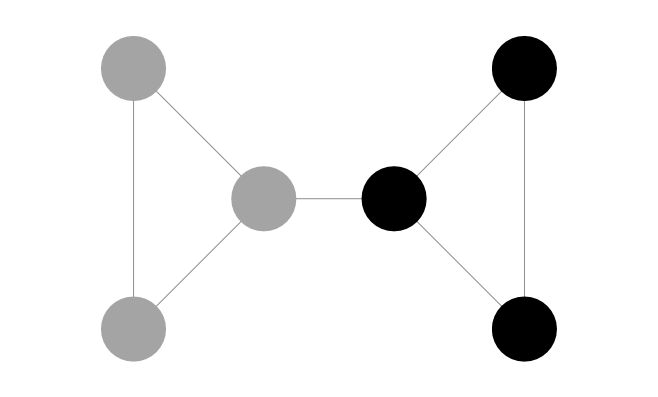}
\caption{A metastable state where social utility is stuck at $-2$.}
\label{graphe_metastable}
\end{center}
\end{figure}





An obvious equilibrium corresponds to the whole economy using a single currency. It is also clearly a social optimum, as no agent has to pay any change cost. However, equation (\ref{utiliteaugmente}) does not imply that social utility will necessarily reach its maximum, which is zero. As figure \ref{graphe_metastable} shows, we can imagine a situation where separated communities appear, within which agents share the same currency, but where several different currencies coexist on the overall economy. Such an equilibrium is clearly not optimal as at least one agent, although suffering losses from the trades with agents from other communities, is nonetheless unable to adopt another currency, as this would entail it to suffer even more losses during some time steps. This contradiction between the individual and collective optima can be found in many economic simple models \cite{hotelling,hernan,pnas}.



\section{Numerical simulations}

\subsection{One community}
\label{sec1commu} 

\begin{figure}
\begin{center}
\includegraphics[width=10cm]{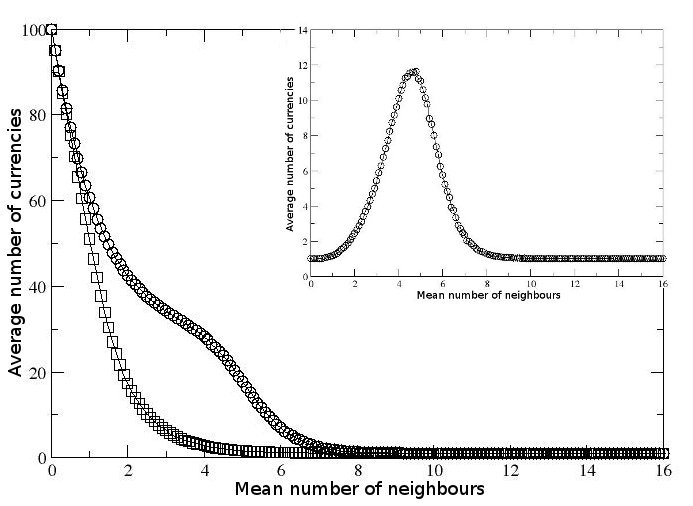}
\caption{At equilibrium for $N=100$, average number of currencies (circles) and of connected components (squares). Insert: average number of currencies per connected component against mean number of neighbours.}
\end{center}
\label{1comu}
\end{figure}

In order to find what kind of equilibria can be reached by the system, we ran $10,000$ simulations of the model. In each simulation, we consider an economy of $N=100$ agents, each beginning with its own currency. Note that two different sources of randomness can influence the results: the topology of the graph and the order in which we pick the agents. Hence, we draw a new distribution of edges for each simulation. 
We then iterate the model, repeating the elementary time steps until equilibrium is reached. 
We count the number of currencies displayed by the overall economy, and calculate its mean over the $10,000$ simulations.

We plot the results for varying link densities $p$, together with the mean number of connected components in the graphs (see figure \ref{1comu}). Indeed, we see that when $p$ is close to $0$, there still are $100$ different currencies in the economy: this was to be expected as 
in this case, many agents have no neighbour at all. They have consequently no other currency to adopt and keep their own each time they are selected to update their situation. 

When the graph cohesion is weak (ie for small values of $p$), we remark that the number of its connected components remains significantly below the average number of currencies remaining at the end of the simulations. 
For $p > 0.1$, which corresponds to an average number of $10$ neighbours per agent, we can reliably consider that there remains only one connected component and that an economy of $100$ agents succeeds in agreeing on a single currency as a mean of exchange for all trades.

\begin{figure}
\begin{center}
\includegraphics[width=8cm]{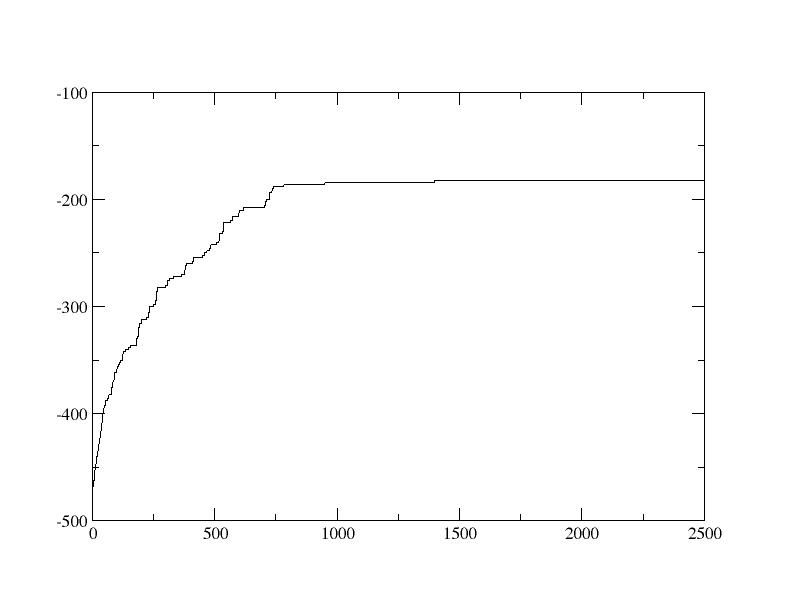}
\caption{Social utility function for $N=100, p=0.05$ and a single connected component. At equilibrium, 14 different currencies are present.}
\label{utilitegenerale}
\end{center}
\end{figure}

We also observe the evolution of social utility through these simulations. Figure \ref{utilitegenerale} shows this evolution in a typical simulation where each agent has on average $5$ neighbours ($p = 0.05$). At equilibrium, which was reached after $1,398$ time steps, social utility culminates at $-182$, which is quite far from the social optimum $0$. Indeed, in this simulation, there remains $14$ different currencies at equilibrium, even if only a single connected component exists. 

We ran the same $10,000$ simulations with two slightly different dynamics. Instead of randomly picking the agents one by one, the second dynamics consisted in computing the optimal choice of each agent in a given state of the economy and then updating their currency choice simultaneously; in the third case, we still picked the agents one by one, but always in the same order. All these variants yielded comparable results.

\subsection{Two communities}
\label{subsec2commu}

\begin{figure}
\begin{tabular}{ccc}
\includegraphics[width = 4cm]{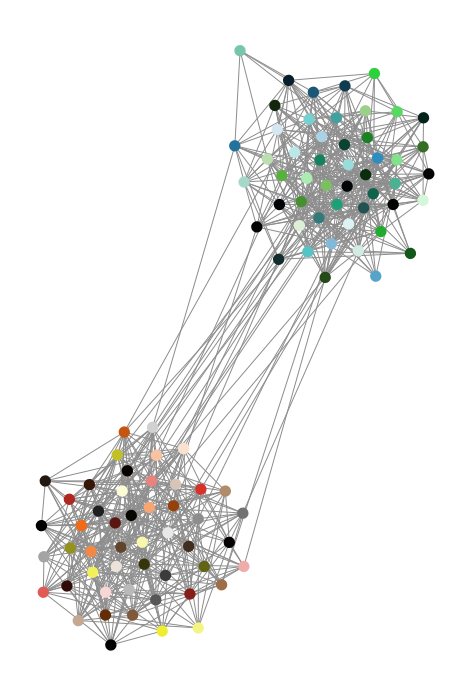} &
\includegraphics[width = 4cm]{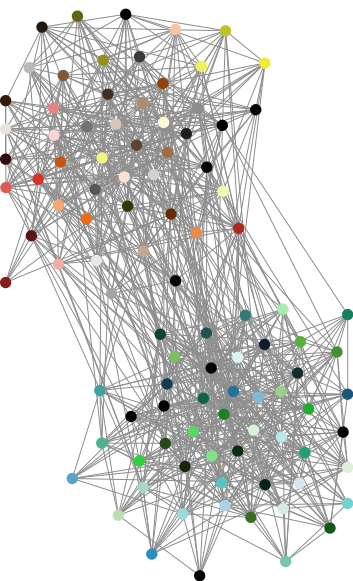} &
\includegraphics[width = 4.2cm]{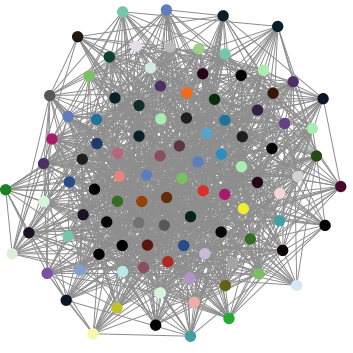}\\
$(a)$ & $(b)$ & $(c)$
\end{tabular}
\caption{Some examples of two-communities graphs with $p_{intra} = 0.3$, for weak [$p_{inter} = 0.01$, figure $(a)$], medium [$p_{inter} = 0.05$, figure $(b)$] and strong [$p_{inter} = 0.2$, figure $(c)$] interconnection.}
\label{graphe2commu}
\end{figure}

Using the same general framework, we now study the case where two different communities composed of $\frac N2$ agents are trading. We consider as a community some pre-defined set of agents sharing a strong intra-connection, while agents belonging to different communities share weaker links. Formally, for an economy consisting of $N$ agents, with $p_{inter}$ and $p_{intra}$ being respectively the probabilities of internal and external links ($p_{inter} < p_{intra}$), we select all pairs of agents $(i,j)$ and create a link with probability $p_{inter}$ if they belong to the same community and with probability $p_{intra}$ otherwise. The mean link density of the overall graph is, for large $N$,
\[p_{av} = \frac 2{N(N-1)} \left(2  \times \frac {\frac N2(\frac N2 -1)}2   p_{intra} + \left(\frac N2\right)^2   p_{inter}\right)
\approx \frac 12(p_{intra} + p_{inter}),\] 

which might be useful to determine the influence of the precise topology on the outcome. In the specific case where $p_{inter} = 0$, there exist two separated communities without {\em intra} links; conversely, if $p_{intra} = p_{inter}$, we actually find the previous case of a single community set on a random graph with uniform link probability $p_{intra}$. Figure \ref{graphe2commu} shows some examples of such graphs for $p_{intra} = 0.3$, from weak inter-connection to strong inter-connection.

Two different cases can be studied: as in the previous one-community case, every agent can begin with its own currency, or each community could be already unified --- which means that its agents all share the same currency.

We begin with the case where each agent is initially given its own currency. 
Just as in the one community case, the question we are concerned with is whether, and on what conditions, the economy as a whole adopts a single currency or if, on the contrary, each community adopts its own currency, leading to a changing cost when trading across them. The result can be expected to depend on the precise system history. If an agent of the first community adopts a currency of the second community and starts spreading its use among its neighbours, the two communities might end using the same currency. But as, on average, an agent shares more links with agents from its own community, she is more likely to adopt a currency from her own community, leading to distinct currencies.

We test the model with $N = 100 \text{ and } p_{intra} = 0.3$. As this is far from the convergence threshold exhibited by experiments with one community, we can figure out what will happen for extreme values of $p_{inter}$. If $p_{inter} = 0$, each community will end with its own currency, as our choice for $p_{intra}$ ensures that a convergence will take place inside each community, and the absence of external links makes impossible the adoption of a currency from the other community. Instead, if $p_{inter} = 0.3 = p_{intra}$, we fall back on the one community case, so that a single currency will spread in the whole economy.

\begin{figure}[h]
\begin{center}
\includegraphics[width=10cm]{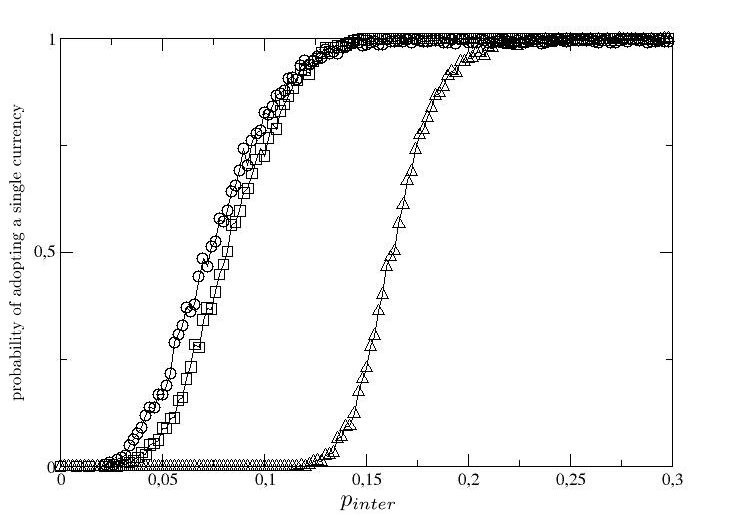}
\caption{Probability that a single currency ermerges on a two-communities graph with $N = 100$ and $p_{intra} = 0.3$. The circles (one agent randomly picked each time step) and the squares (all agents changing their choice simultaneously) represent the case when each agent begins with its own currency as in the one-community case. The triangles shows the transition when each community had previously been unified on a single currency.}
\label{2commu}
\end{center}
\end{figure}

In order to see how the transition takes place, we run $10,000$ simulations for several values of $p_{inter}$ and plot the probability that a single currency is adopted in the whole economy at equilibrium.
For each simulation, we draw a new random graph as before. The results are represented by the circles on figure \ref{2commu}. The curves agree with our predictions: for low values of $p_{inter}$, each community uses a different currency; but as $p_{inter}$ increases, the economy tends to adopt a single currency. Here again, we explore what happens when all the agents update their choice simultaneously instead of one by one (represented by the squares in figure \ref{2commu}), but the two curves are almost identical.

However, this result must be compared to the one-community case studied before: even for $p_{inter} = 0$, the link density of the overall graph is still $\frac 12p_{intra} = 0.15$. For an equivalent link density, we saw in section \ref{sec1commu} that when the graph was purely random, with such a mean link density the economy almost always united on a single currency.

We now turn to the case where each community has already been unified on a single currency. 
The problem is then to know whether the attractiveness of the other community is strong enough to counter the intra-communitarian bonds.

We ran again $10,000$ simulations for an economy with $N=100$ and $p_{intra} = 0.3$, and plotted the results for varying $p_{inter}$ (represented by the triangles on figure \ref{2commu}). As was to be expected, the transition from two to one currency needs a much higher value of $p_{inter}$ to happen. Indeed, the the intra-communitarian bonds are \textit{already effective} when agents are offered to switch currency; whereas in the previous case, the bonds required to be activated by the --- possible --- adoption of a community-dependent currency to become real constraints, and hence was left open the possibility that the agents chose a currency used in both communities.

\section{Exact results for two large pre-unified communities}

We now show rigorously that for two very large communities ($p_{inter} < p_{intra}, N\to\infty$), a transition to a single currency never happens.  Intuitively, the transition may be triggered by a single agent having more external than internal neighbors, which is unlikely but not impossible when $p_{inter} < p_{intra}$, because of random fluctuations. One expects that fluctuations in the links' distributions become smaller as N increases, leading to a more stable advantage of inner links. However, since the number of nodes increases, the probability that at least one node has more external than internal links may also increase, leading to a cascade. Our calculations prove that the decrease in the fluctuations is stronger than the increase in the number of nodes. More precisely, if $\mathcal P_N$ is the probability that, for a given two-community graph with $N$ agents, there exists a single-currency equilibrium, we demonstrate that, when $p_{inter} > p_{intra}$, $\mathcal P_N \to 0$ as $N\to\infty$.

An obvious necessary condition for the existence of such an equilibrium is that there exists at least one agent which might change its currency. We thus assume $p_{inter} < p_{intra}$ and begin by studying the probability, for a given agent, to have less neighbours in its own community than in the other.

For a given $N$, we randomly choose one of the agents and define $X_{a,N}$ as its number of \textit{intra}communitarian links and $X_{r,N}$ as its number of \textit{inter}communitarian links. For random graphs, $X_{a,N}$ and $X_{r,N}$ are two independent random variables following binomial laws with $N$ number of trials and respective parameters $p_{intra}$ and $p_{inter}$. Let $\phi_a^N$ and $\phi_r^N$ respectively denote their cumulative distribution function.

We thus study the probability
\[
\begin{split}
P(X_{a,N} \leq X_{r,N}) 	& = \sum_{k=0}^{\infty} P(X_{r,N}=k \text{ and } X_{a,N} \leq k) \\
					& = \sum_{k=0}^{\infty} P(X_{r,N}=k)P(X_{a,N} \leq k) \\
					& = \sum_{k=0}^{\infty} P(X_{r,N}=k)\phi_a^N(k) .
\end{split}
\]

Let $p_{av} = \frac{p_{intra} + p_{inter} }2$; as $p_{inter} < p_{intra}$, then $p_{inter} < p_{av} < p_{intra}$. For each $N$, we define $k_N = \lfloor p_{av} N \rfloor$ the floor of $p_{av} N$. We obtain:

\begin{eqnarray}
P(X_{a,N} \leq X_{r,N}) 	& = 		& \sum_{k=0}^{k_N} P(X_{r,N}=k)\underbracket{\phi_a^N(k)}_{\leq \phi_a^N(k_N)}
										 + \sum_{k=k_N}^{\infty} P(X_{r,N}=k)\underbracket{\phi_a^N(k)}_{\leq 1} \nonumber\\
					& \leq 	& \phi_a^N(k_N)\sum_{k=0}^{k_N} P(X_{r,N}=k)
										 + \sum_{k=k_N}^{\infty} P(X_{r,N}=k) \nonumber\\
					& \leq	& \phi_a^N(k_N) + (1 - \phi_r^N(k_N)) \label{somme}
\end{eqnarray}

We will now show that 
$
\left\{
\begin{array}{c}
\phi_a^N(k_N) \underset{N\to +\infty}\longrightarrow 0 \\
\text{and} \\
1-\phi_r^N(k_N) \underset{N\to +\infty}\longrightarrow 0
\end{array}
\right.
$
, and show that these limits follow at least a geometrical decay. 

By definition:
\[ \phi_a^N(k_N) = \sum_{k=0}^{k_N}\binom Nk p^k (1-p)^{N-k},\]
where we use $p$ instead of $p_{intra}$ to simplify notations. Moreover, if $ k\leq p_{av} N$, $
\binom Nk 		 = \frac N{N-k}\binom {N-1}k \leq \frac 1{1-p_{av}} \binom {N-1}k.$ Hence:
%
%
\begin{eqnarray}
\phi_a^N(k_N)
     & \leq 	&	\sum_{k=0}^{k_N}\frac 1{1-p_{av}}\binom {N-1}k p^k (1-p)^{N-k} \nonumber\\
      & \leq	&	... \nonumber\\
      & \leq	&	\sum_{k=0}^{k_N}\left(\frac 1{1-p_{av}}\right)^{N-k_N}\binom {k_N}k p^k (1-p)^{N-k} \nonumber\\
      & =	&	\left(\frac {1-p}{1-p_{av}}\right)^{N-k_N}\ \underbracket{\sum_{k=0}^{k_N}\binom {k_N}k p^k (1-p)^{k_N-k}}_{= 1} \nonumber\\
      & \leq	&	\rho_a^N \label{rho_a}
\end{eqnarray}
where $\rho_a = \left(\frac {1-p_{intra}}{1-p_{av}}\right)^{1-p_{av}} < 1$.

Similarly, if we now use $p$ for $p_{inter}$, we get:
\[
\begin{split}
1 - \phi_r^N(k_N)	\ \ 	& =\ \  1 - \sum_{k=0}^{k_N}\binom Nk p^k (1-p)^{N-k} \\
				\ \ 	& =\ \  \sum_{k=k_N + 1}^N\binom Nk p^k (1-p)^{N-k} \\
				\ \ 	& =\ \  \sum_{k'=0}^{N-k_N - 1}\binom N{k'} p^{N-k'} (1-p)^{k'}
\end{split}
\]

Moreover, if $k\leq N-k_N-1=N-\lfloor p_{av} N\rfloor-1$, then $k\leq (1-p_{av})N$ and $\binom Nk\leq\frac 1p_{av}\binom {N-1}k$. From this point,

\begin{eqnarray}
1 - \phi_r^N(k_N)
	& \leq	& \sum_{k'=0}^{N-k_N-1}\left(\frac 1p_{av}\right)^{k_N+1}\binom {N-k_N-1}{k'} p^{N-k'} (1-p)^{k'}\nonumber\\
	& = 		& \left(\frac pp_{av}\right)^{k_N+1}\ \sum_{k'=0}^{N-k_N-1}\binom {N-k_N-1}{k'} p^{N-k_N-1-k'} (1-p)^{k'}\nonumber\\
	& \leq\ 	& \rho_r^N \label{rho_r}
\end{eqnarray}
where $\rho_r = \left(\frac {p_{inter}}{p_{av}}\right)^{p_{av}}<1$.

Putting equations (\ref{somme}), (\ref{rho_a}) and (\ref{rho_r}) together, we get

\begin{equation}
P(X_{a,N} \leq X_{r,N}) \leq \rho_a^N + \rho_r^N,
\label{majorationgeometrique}
\end{equation}
which means that the probability for a single agent to have more neighbours in the other community than in its own geometrically decreases to $0$ as $N$ tends to infinity.

For $i = 1,...,N$, let us now call $X_{a,N}^{(i)}$ and $X_{r,N}^{(i)}$ the random variables representing, respectively, the number of \textit{intra}communitarian and of \textit{inter}commu\-ni\-ta\-rian neighbours of agent $i$. Recall that $\mathcal P_N$ is the probability that, for a given two-community graph with $N$ agents, there exists a single-currency equilibrium. A necessary condition is that there exists some agent in the initial configuration which may change its currency. Using equation (\ref{majorationgeometrique}), we can then write:
\begin{eqnarray}
\mathcal P_N 	
& \leq	& P\left(X_{a,N}^{(1)} \leq X_{r,N}^{(1)}\ \text{ or \ ... \ or }\ X_{a,N}^{(N)} \leq X_{r,N}^{(N)}\right) \nonumber\\
& \leq	& P\left(X_{a,N}^{(1)} \leq X_{r,N}^{(1)}\right) + ... + P\left(X_{a,N}^{(N)} \leq X_{r,N}^{(N)}\right) \nonumber\\
& =		& NP\left(X_{a,N} \leq X_{r,N}\right)\nonumber\\
&\leq	& N\left(\rho_a^N + \rho_r^N\right)\label{majorationfinale}\nonumber
\end{eqnarray}
which shows that $\mathcal P_N \underset{N\to\infty}\longrightarrow0$.




\section{Agents heterogeneity}

Until now, we have supposed that every agent has the same influence on its neighbours: the loss due to using different currencies is $1$. We now relax this assumption by introducing heterogeneity among the agents. We assume that each agent $j$ has a weight $\rho_j$; when an agent $i$ trades with $j$ using a currency $s_i\neq s_j$, it now has to afford a trading cost $\rho_j$, so that we can define a new individual utility function for the agent $i$:

\[\tilde{U_i} = - \sum_{j\in \mathcal V(i)}\rho_j(1 - \delta_{s_i}^{s_j}) \]

We first set $\rho_i = deg(i)$ where $deg(i)$ is the degree of agent $i$ in the graph of commercial links
. We then run the same simulations as before (for one community, and for two communities either unified or not) and find no qualitative difference in the transition curve. However, introducing weights has an important impact on \textit{which} currency will eventually be chosen: is the case when each agent begins with its own currency, the average degree (hence the average weight) of the agent whose currency will be adopted is much higher than when the agents are unweighted. 

We then attribute randomly the $\rho_i$, independently of agent's degree, but keeping the same Poisson distribution as in the random graph. We obtain the same results: the transition curves for currency adoption remain unchanged, but the average weight of the first owner of the adopted currency is much higher than the average weight.

\section{Conclusion}
We have studied a simplistic model showing coexistence of several currencies, even in presence of increasing returns to adoption. Agents exchange only with a limited number of neighbors, through local exogenous commercial links, and seek to minimize their transaction costs by adopting the most common currency among these. The main interest of our work is to provide a model that is at the same time very simple in its structure (exchange network and transaction mechanism) but is able to recover as possible equilibria both the existence of a single currency or several of them, while most previous models only found the single currency equilibrium \cite{jones,greenaway,sethi,lapavitsas,gangotena,yasutomi,gorski10,gorski15}. This last point reminds work by Brian Arthur \cite{arthur}, where increasing returns lead to the existence of multiple equilibria. In further explorations, it would be interesting to investigate to which extent these fractionalized stationary states are robust to noise in the decision process, for example by introducing a trembling hand, the possibility for an agent to choose, with a small probability a currency that is not the most common among his neighbors.

\bibliographystyle{unsrt}
\bibliography{refs}{}

\end{document}